\documentclass[aps,twocolumn,superscriptaddress]{revtex4-2} 

\usepackage[utf8]{inputenc}
\usepackage[dvips]{graphicx}
\usepackage{xcolor}
\usepackage{braket}
\usepackage{bm,amsmath,amsfonts,amsthm,amssymb,mathtools}
\usepackage[bookmarksnumbered,pdfpagelabels=true,plainpages=false,colorlinks=true,linkcolor=blue,citecolor=red,urlcolor=blue]{hyperref}
\usepackage{natbib}
\usepackage[export]{adjustbox}
\usepackage[capitalise]{cleveref}

\usepackage{comment}

\def\ri{\mathrm i}
\def\re{\mathrm e}

\def\pdag{{\vphantom\dag}}

\def\tr{{\rm tr}}

\begin{document}

\title{Quantum Magnetic Skyrmion Operator}

\author{Andreas Haller}
\affiliation{Department of Physics and Materials Science, University of Luxembourg, 1511 Luxembourg, Luxembourg}

\author{Sebasti{\'a}n A. D{\'i}az}
\affiliation{Department of Physics, University of Konstanz, 78457 Konstanz, Germany}

\author{Wolfgang Belzig}
\affiliation{Department of Physics, University of Konstanz, 78457 Konstanz, Germany}

\author{Thomas L. Schmidt}
\affiliation{Department of Physics and Materials Science, University of Luxembourg, 1511 Luxembourg, Luxembourg}

\date{\today}

\begin{abstract}
    We propose a variational wave function to represent quantum skyrmions as bosonic operators.
    The operator faithfully reproduces two fundamental features of quantum skyrmions: their classical magnetic order and a ``quantum cloud'' of local spin-flip excitations.
    Using exact numerical simulations of the ground states of a 2D chiral magnetic model, we find two regions in the single-skyrmion state diagram distinguished by their leading quantum corrections.
    We use matrix product state simulations of the adiabatic braiding of two skyrmions to verify that the operator representation of skyrmions is valid at large inter-skyrmion distances.
    Our work demonstrates that skyrmions can be approximately coarse-grained and represented by bosonic quasiparticles, which paves the way toward a field theory of many-skyrmion quantum phases and, unlike other approaches, incorporates the microscopic quantum fluctuations of individual skyrmions.
\end{abstract}

\maketitle

{\it Introduction.--}
Magnetic skyrmions are topologically nontrivial spin configurations that often arise in non-centrosymmetric magnetic materials~\cite{Bogdanov1989,Bogdanov1994,Rossler2006,Neubauer2009} and have mostly admitted a description in terms of classical magnetic moments~\cite{EverschorSitte2018}.
However, by now several materials have been found where the sizes of individual skyrmions can be on the order of the spacing of the underlying atomic lattice~\cite{Heinze2011,Nagaosa2013}.
This makes it necessary to explore their properties based on the underlying quantum mechanical models. 
Quantum fluctuations, superposition states, and entanglement become unavoidable and fascinating features of quantum magnetic skyrmions that remain largely unexplored despite their relevance for the emerging proposals of skyrmion-based quantum technologies.
Such applications include the potential use of skyrmions as qubits in frustrated magnets~\cite{Psaroudaki2021,Xia2023} and as mobile impurities to implement topological quantum computing in magnet-superconductor heterostructures~\cite{Nothhelfer2022,Diaz2021b}.
The fundamental quantum features of magnetic skyrmions eventually render a classical description unsuitable, hence calling for a new and inherently quantum mechanical theoretical framework.

Quantum magnetic skyrmions emerge, for instance, as the many-body ground state of ferromagnetic spin-$1/2$ Heisenberg models with either Dzyaloshinskii-Moriya interactions or frustration due to antiferromagnetic next-nearest neighbor couplings.
Since an exact numerical diagonalization of such spin Hamiltonians becomes prohibitive for a large number of lattice sites, the theoretical description of quantum magnetic skyrmions has relied mainly on alternative approaches.
Near the classical limit, collective coordinates together with path integral techniques have been employed to model quantum skyrmion nucleation~\cite{Vlasov2020,Diaz2021} and quantum skyrmion dynamics~\cite{Psaroudaki2017,Ochoa2019}.
These approaches, unfortunately, are inadequate to model systems of interest for quantum technologies: quantum skyrmions in large numbers and in the deep quantum regime.
Here, quantitative predictions for few-skyrmion systems can be made based on exact diagonalization~\cite{Lohani2019,Sotnikov2021,Sotnikov2023,Vijayan2023,Mazurenko2023}, neural network states~\cite{Joshi2023,Joshi2024_preprint}, dynamical mean-field simulations~\cite{Peters2023}, and density matrix renormalization group simulations~\cite{Haller2022,Zhao2024}.

In this work, we develop a theoretical framework that makes it possible to describe this regime.
It is based on the construction of creation and annihilation operators of quantum skyrmions and thus treats magnetic skyrmions as quantum quasiparticles.
This compact and elegant theoretical framework captures the quantum skyrmion's defining features: a classical magnetic order with quantum fluctuations.
Constructing this operator allows us to clarify the role played by quantum fluctuations and to determine the range of validity of the semiclassical approximation.
The quantum magnetic skyrmion operator is a natural and computationally efficient starting point to model skyrmion-skyrmion interactions and quantum many-body skyrmion phenomena.

\begin{figure}
    \centering
    \includegraphics{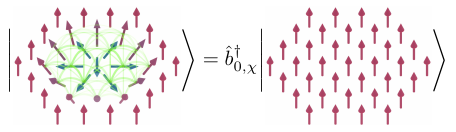}
    \caption{Creation of a quantum magnetic skyrmion.
    The quantum magnetic skyrmion operator $\hat b^\dag_{0,\chi}$ acts on the field-polarized vacuum state to create a quantum magnetic skyrmion.
    By construction, the operator captures the fundamental features: classical magnetic order (arrows) and quantum fluctuations visualized by pairwise spin entanglement (green lines).}
\end{figure}

{\it Quantum Skyrmion Operator.--}
The proposed construction of the quantum skyrmion operator relies on the assumption that a magnetic quantum skyrmion has a typical magnetic order in the local spin expectation values.
This includes cases where the norm of the vector of spin expectation values is not quantized to the quantum spin $s$, and we seek to faithfully capture the quantum fluctuations around the magnetic order.
In the following, we focus on the special case of $s=1/2$, but our construction admits a natural extension to any spin $s$.
We start by defining a product of local spin rotations
\begin{align}
    \hat R_{\overline0}(\bm\theta,\bm\phi) = \prod_{j\neq 0}\re^{-\ri \phi_j \hat S_{j,z}}\re^{-\ri \theta_j \hat S_{j,y}},
    \label{eq:rnot0}
\end{align}
where the subscript $\overline 0$ indicates that the site $j = 0$ is excluded from the operator product. For suitably chosen angles $\phi_j$ and $\theta_j$, we may write a classical magnetic skyrmion, which is represented by a product state in real space, in the form
\begin{align}
    \ket{\Psi_c(\bm\theta,\bm\phi)} = \hat R_{\overline0}(\bm\theta,\bm\phi)\hat S^{-}_{0}\ket{0}
    \label{eq:classical_skyrmion_creation_operator}
\end{align}
where the fully spin-polarized state $\ket{0}=\ket{\Uparrow} = \otimes_i\ket{\uparrow}_i$ acts as the vacuum state and the spin at site $j = 0$ is flipped with respect to the polarized environment.
We can therefore interpret $\hat b^\dag_{0,0} = \hat R_{\overline 0}\hat S^{-}_{0}$ as a creation operator for the classical skyrmion $\ket{\Psi_c}$ from the vacuum $\ket{0}$.

Since $[\hat R_{\overline0},\hat S^-_{0}]=0$, the local commutation relations of $\hat b^\dag_{0,0}$ are given by $\{\hat b^\pdag_{0,0},\hat b^\dag_{0,0}\}=1$ and $[\hat b^\pdag_{0,0},\hat b^\dag_{0,0}]=2\hat S_{0,z}$.
We can analogously define creation operators for all sites of the lattice by $\hat b^\dag_{i,0} = \hat R_{\overline i} \hat S^{-}_{i}$ by shifting all operators from site $0$ to site $i$.
Clearly, the commutator of $\hat b_{i,0}$ and $\hat b^\dag_{j,0}$ (or $\hat b_{j,0}$) vanishes if sites $i$ and $j$ are distant enough such that the local rotations $\phi_j, \theta_j$ contributing to $\hat R_{\overline i}$ and $\hat R_{\overline j}$ act on disjoint sets of sites.

To incorporate quantum fluctuations around the classical order, we define the following quantum skyrmion creation operator,
\begin{align}
    \hat b^\dag_{i,\chi} 
    &= \frac{\hat R_{\overline i}}{\sqrt{\sum_{k=0}^\chi|w_k|^2}} \sum_{k=0}^\chi w_k \left(\prod_{j} \left(\hat S^-_j\right)^{n_{k,j}} \right),
    \label{eq:quantum_skyrmion_operator}
\end{align}
which is the central object of this paper.
The complex scalar weights $\{w_0, \ldots, w_\chi\}$ are sorted by descending absolute value, $|w_i| \geq |w_{i+1}|$, $\bm n_k\in \{0,1\}^{\otimes N}$ is a vector representing the number of ladder operators in the product of spin flips, and $\chi$ is a so far arbitrary integer which is bounded by the dimension of the Hilbert space.
\Cref{eq:quantum_skyrmion_operator} is motivated by the observation that magnetically ordered quantum skyrmion ground states centered at position $i$ are in general surrounded by a ``cloud'' of spin-flip fluctuations \cite{Haller2022,Joshi2023,Joshi2024_preprint}.
For semiclassical skyrmion profiles one has $\bm n_0$ with $n_{0,i}=1$ and $n_{0,j\neq i}=0$ associated with the leading contribution $|w_0| > |w_{k\neq 0}|$ from the classical order.
The precise form of other vectors $\bm n_{k\neq0}$, corresponding to the ``quantum cloud'', remain undetermined and are model-dependent.
It is straightforward to recognize that the local commutation relation of quantum skyrmion operators is approximately given by
\begin{align}
    \left[\hat b_{i,\chi}, \hat b^\dag_{i,\chi}\right] \approx \frac{|w_0|^2}{\sum_k |w_k|^2}2\hat S_{i,z}
    ,\quad
    \left\{\hat b_{i,\chi}, \hat b^\dag_{i,\chi}\right\} \approx \frac{|w_0|^2}{\sum_k |w_k|^2}
    ,
    \label{eq:quantum_skyrmion_operator_commutation_relations}
\end{align}
together with
\begin{align}
    \left[\hat b_{i,\chi}, \hat b^\dag_{j,\chi}\right] = 0
    \text{ for }
    |\bm R_j-\bm R_i| > r_c
    \label{eq:quantum_skyrmion_operator_commutation_relations_nonlocal}
\end{align}
in which $r_c$ denotes the radius of the skyrmion profile.
In total, one can therefore interpret the quantum skyrmion operator as a bosonic entity, but with a constraint on the local Hilbert space dictated by \cref{eq:quantum_skyrmion_operator_commutation_relations}.
Using matrix product state simulations, it is possible to extract the phase after an adiabatic exchange of a system containing two quantum skyrmions, and we find that the exchange phase is compatible with \cref{eq:quantum_skyrmion_operator_commutation_relations_nonlocal} (see Supplemental Material~\cite{SM}).
Moreover, the local ``hard-core'' constraint of \cref{eq:quantum_skyrmion_operator_commutation_relations} agrees with the numerical observation that many-skyrmion ground states are composed of skyrmion crystals and liquids~\cite{Haller2022}.
In principle, an operator of the form \cref{eq:quantum_skyrmion_operator} can be used to construct a complete basis of the Hilbert space (see~\cite{SM}).
In the following, we will refer to this basis as the rotated Fock basis.
However, the purpose of the remaining paper is to illustrate for a particular magnetic model that the ratios of the scalar weights $w_i/w_0$ make it possible to truncate the sum to values $\chi$ much smaller than the dimension of the Hilbert space, while low-lying single-particle quantum skyrmion states are still well approximated by $\hat b^\dag_{i,\chi}\ket{0}$.

\begin{figure*}[ht]
    \includegraphics[scale=0.868]{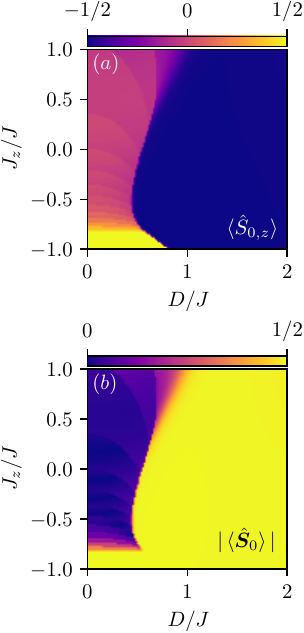}
    \hfill
    \includegraphics[scale=0.868]{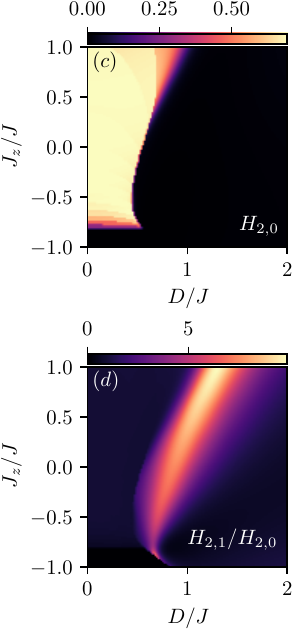}
    \hfill
    \includegraphics[scale=0.868]{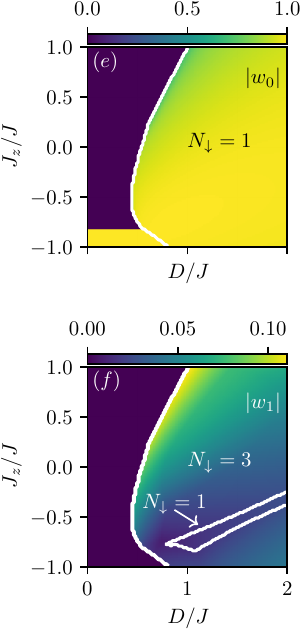}
    \hfill
    \includegraphics[scale=0.868]{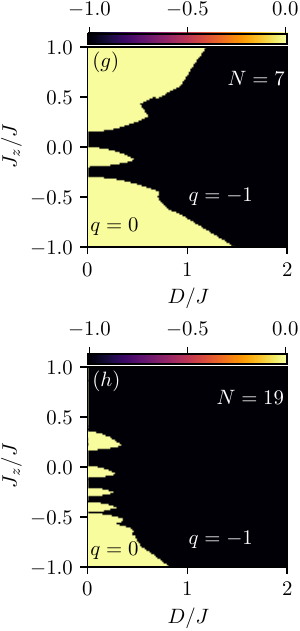}
    \caption{
    Quantum magnetic skyrmions and other ground states stabilized in a finite-sized ``quantum flake.''
    The quantum flake -- a hexagonal region of the lattice of $N$ spins with $s = 1/2$ -- is part of a chiral magnet embedded in a $m_z = +1/2$ polarized classical environment.
    The Hamiltonian~\cref{eq:boundary_conditions} models the coupling to the classical environment and \cref{eq:hamiltonian} governs the chiral magnet, in which we use a vanishing Zeeman field $B = 0$ and $J_x = J_y = -J$ ($J>0$) for the exchange couplings.
    Varying the DMI amplitude $D$ and the exchange coupling $J_z$, we use exact diagonalization to compute ground state quantities for $N = 19$ sites:
    $(a)$ expectation value of the central spin $\braket{\hat S_{0,z}}$;
    $(b)$ norm of the central spin $|\braket{\hat{\bm S}_{0}}|$;
    $(c)$ center second Rényi entropy $H_{2,0}$;
    $(d)$ ratio between off-center and center second Rényi entropies $H_{2,1}/H_{2,0}$;
    probability amplitude magnitudes of the $(e)$ classical magnetic order $|w_0|$ and $(f)$ the six leading quantum corrections with equal amplitudes $|w_1|=|w_2|=\dots=|w_6|$, defined in \cref{eq:skyrmion_wave_function}, from the ground state expanded in the rotated Fock basis; and $(h)$ topological charge of the ground state magnetization $q$ (defined in~\cite{SM}).
    In panel $(g)$, we depict $q$ for an $N = 7$ quantum flake.
    Possible ground states: quantum skyrmion (blue region in $(a)$ where $\braket{\hat S_{0,z}} \approx -1/2$), field-polarized state (yellow region in $(a)$ where $\braket{\hat S_{0,z}} = +1/2$), and vortex state [pink region in $(a)$ where $\braket{\hat S_{0,z}} \approx 0$ with large quantum fluctuations indicated in $(c)$].
    Contour lines and $N_\downarrow=\sum_j n_{k,j}$ in $(e)$ and $(f)$ indicate, respectively, the region and the corresponding number of down spins.
    In \cref{fig:sub_regions}, we show that the $N_\downarrow=3$ region can be divided into further sub-regions, distinguished by the distribution of spin flips in the quantum corrections.
    Probability amplitudes $w_0$ and $w_1$ are the weights for constructing the quantum skyrmion operator.
    }
    \label{fig:state_diagram}
\end{figure*}

{\it Quantum Skyrmions in Chiral Magnets.--}
We investigate a minimal model of a two-dimensional chiral magnet
\begin{align}
    \hat H
    &=
    \sum_{\braket{ij}}\left(\sum_{\alpha = x,y,z} J_{\alpha}\hat{S}_{i,\alpha} \hat{S}_{j,\alpha} + \bm D_{ij}\cdot\left(\hat{\bm S}_i\times\hat{\bm S}_j\right)\right)
    \nonumber
    \\
    &
    \qquad
    +
    \sum_i\bm B\cdot\hat{\bm S}_i
    \label{eq:hamiltonian}
\end{align}
where $\braket{ij}$ denotes a sum over pairs of neighboring sites (each pair is only summed over once).
Unless otherwise mentioned, the operators $\hat{\bm S}_i$ are spin $s=1/2$ operators and we assume an interfacial Dzyaloshinskii–Moriya interaction (DMI) with $\bm D_{ij} = D \hat e_z\times\bm r_{ij}/a$ where $a$ is the lattice constant and $\bm r_{ij}$ the distance vector between two neighboring sites.

It has been shown that a system described by \cref{eq:hamiltonian} hosts a variety of quantum skyrmion ground states~\cite{Sotnikov2021,Siegl2022,Vijayan2023,Mazurenko2023,Joshi2023,Salvati2024}.
For small systems, finite-size effects are not negligible and the properties of the low-energy states can change dramatically with the choice of boundary conditions: 
open boundary conditions lead to helical magnetic spirals, field-polarized states, semiclassical skyrmions, liquids, and lattices; periodic boundary conditions can give rise to a symmetric and translation invariant ground state~\cite{Sotnikov2021}; while a quantum flake embedded in a classical spin field suppresses helical states and removes the chiral surface twist on the system boundary~\cite{Siegl2022}.
To avoid such boundary effects, we focus on a quantum flake embedded in a fully polarized classical spin field.
The quantum spins are located on a finite subset $\mathcal Q$ of the full lattice $\mathcal S$, i.e., $\mathcal Q = \{\bm R_i, i=0, \ldots, N-1\} \subset \mathcal S$, and are described by \cref{eq:hamiltonian}.
Moreover, we add the following term to account for the coupling to the environment:
\begin{align}
    \hat H_{\mathcal Q\mathcal C} = \sum_{\braket{ij}}\left(\sum_{\alpha = x,y,z} J_{\alpha}\hat{S}_{i,\alpha} {m}_{j,\alpha} + \bm D_{ij}\cdot\left(\hat{\bm S}_i\times{\bm m}_j\right)\right)
    \label{eq:boundary_conditions}
\end{align}
where the sites $j$ are in the complement $\mathcal C = \mathcal S \backslash \mathcal Q$ and $\bm m_j = (m_{j,x}, m_{j,y}, m_{j,z})$ is a classical spin with length $s$.

For computational convenience, we consider finite triangular lattices with hexagonal $C_6$ symmetry.
We call the $N=7$ flake a single-shell and the $N=19$ flake a two-shell system, which allows us to investigate by exact diagonalization small $C_6$ symmetric quantum skyrmions which have a maximal radius $r_c\approx 3a$.
Some quantum properties of these states have already been investigated numerically in larger systems with open boundary conditions~\cite{Haller2022,Joshi2023}, but quantifying the amount of quantum fluctuations, and how these fluctuations are affected by microscopic parameters, remains largely unexplored.

In the next section, we examine the skyrmion ground states $\ket\Psi$ obtained by exact diagonalization and perform a tomography in the rotated Fock basis which allows us to argue that the creation operator~\eqref{eq:quantum_skyrmion_operator} indeed satisfies $\ket\Psi \approx \hat b^\dag_{0,\chi}\ket0$, where the quality of the approximation is controlled by $\chi$.

\begin{figure}[ht]
    \centering
    \includegraphics{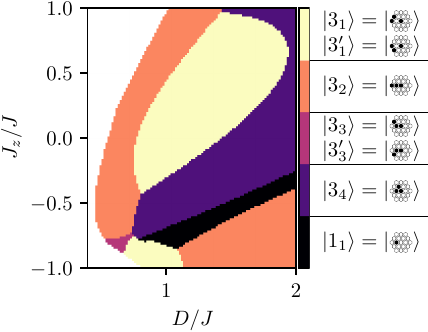}
    \caption{Regions of different quantum skyrmions, distinguished by their leading quantum corrections (second terms in \cref{eq:skyrmion_wave_function}). The Fock states contributing to these corrections are equal up to $\hat C_6$ rotations, but differ in the distribution of the down spins, indicated by the graphical notation. States in sub-region $2$ and $4$ do not have mirror-symmetric partner states, i.e. $\ket{3_2'}=\ket{3_4'}=0$. The black region corresponds to the $N_\downarrow=1$ region of \cref{fig:state_diagram}~$(f)$ with states $\ket{\Psi_{\rm II}}$, and the other regions are sub-regions of $N_\downarrow=3$.}
    \label{fig:sub_regions}
\end{figure}

{\it Tomography of Quantum Skyrmions.--}
Using exact diagonalization, we obtain the ground state diagram of \cref{eq:hamiltonian} subject to the boundary conditions imposed by adding \cref{eq:boundary_conditions} and present our findings in \cref{fig:state_diagram}.
The state diagram we obtain here is qualitatively similar to that presented in a previous work~\cite{Siegl2022}, but it deserves further discussion.
Our exact diagonalization simulations are performed with $19$ sites on a triangular lattice whereas the previous work used $9$ sites on a square lattice.
Qualitatively, local spin expectation values allow a clear distinction between three regions: (i) a region hosting a field-polarized classical ground state indicated by a quantized spin expectation value norm $s = |\braket{\hat{\bm S}_{0}}|$ in the yellow regime of \cref{fig:state_diagram}~$(a)$, (ii) a vortex region with large quantum fluctuations signaled by vanishing center spin expectation value $|\braket{\hat{\bm S}_0}|\approx 0$ and center Rényi entropy $H_{2,0}\approx\ln2$ displayed in \cref{fig:state_diagram}~$(b)$ and $(c)$, and (iii) a quantum skyrmion regime which is indicated by the contour lines in panels $(e)$ and $(f)$.
The quantum skyrmion region is also clearly indicated by an opposite magnetization order compared to the field-polarized environment, which is marked by the dark blue region in \cref{fig:state_diagram}~$(a)$, and a finite spin expectation value norm presented in panel $(b)$.
The field-polarized and quantum skyrmion states can further be distinguished by investigating the ratio between off-center and center Rényi entropy, defined by
\begin{align}
    H_\alpha(\rho_A) = \frac1{1-\alpha}\tr\ln(\rho^\alpha_A)
    ,
    \quad
    \rho_A = \tr_{A^C}\ket\Psi\bra\Psi
    ,
    \label{eq:Renyi_entropy}
\end{align}
where $\rho_A$ is the reduced density matrix of a bipartition $A\subseteq\mathcal Q$ and the trace is performed over the local basis of the complement $A^C$.
In \cref{fig:state_diagram}, panels $(c)$ and $(d)$, we show the second Rényi entropies for a single-spin bipartition $A_i=\{\bm R_i\}$, i.e., $H_{2,i}=H_2(\rho_{A_i})$.
In $(c)$, the single spin is at the center of the flake, which we denote by $A_0=\{\bm R_0\}$ and $H_{2,0}$, whereas $(d)$ presents the ratio between the off-center Rényi entropies $H_{2,1}/H_{2,0}$, where $A_1=\{\bm R_1\}$ denotes the bipartition choice of the off-center spin, $|\bm R_1 - \bm R_0|=a$.
Due to the six-fold rotation symmetry of the flake, it is irrelevant which spin of the first shell is chosen.
The ratio $H_{2,1}/H_{2,0}$, displayed in \cref{fig:state_diagram}~$(d)$, indicates that the off-center spins of the skyrmion texture can be more entangled with the rest of the quantum system than the center spin.
This is compatible with the entanglement properties found for skyrmions of larger systems~\cite{Haller2022,Joshi2023,Bhowmick2023_inprep}.
We want to emphasize here that the field-polarized states are product states, indicated by vanishing Rényi entropy for every spin, whereas quantum skyrmion states display small but non-vanishing Rényi entropy.
Although the state boundaries between vortex and magnetically ordered states (yellow and blue regions of \cref{fig:state_diagram}~$(b)$) are slightly shifted upon varying the system geometry, we find that the qualitative features of the magnetically ordered states in the yellow region of panel $(b)$ remain the same, and therefore finite size effects are marginal.
In contrast, the vortex states of the blue region in \cref{fig:state_diagram}~$(b)$ are characterized by near-maximal center site Rényi entropy (displayed in \cref{fig:state_diagram}~$(c)$) and almost vanishing spin expectation norm.
Previous work identified some states in the $q=-1$ lobes of panels \cref{fig:state_diagram}~$(g)$ and $(h)$ of the vortex region as quantum skyrmions due to the possibility of defining a quantized topological charge~\cite{Siegl2022}.
However, the form of the magnetic profile and the resulting topological charge of the regions with small non-vanishing spin expectations depend strongly on the geometry of the quantum system.
We thus focus on a detailed investigation of the unambiguously ordered quantum skyrmion states, for which the quantum skyrmion operator is most relevant.
Using the angles associated with the local magnetic order, we expand the wave function in a rotated Fock basis~\cite{SM}, and sort the weights by absolute magnitude.
These weights can be extracted from the state obtained by exact diagonalization, and are shown in~\cref{fig:state_diagram}~$(e, f)$.
Depending on the parameters of the system, we find that the ground state wave functions assume two qualitatively different forms, which we denote as $\ket{\Psi_{{\rm I},i}}$ and $\ket{\Psi_{\rm II}}$, i.e.,
\begin{align}
    &\ket{\Psi_{{\rm I},i}} = \hat R_{\overline0}
        \left(
            w_0
            \ket{1_0}
            +
            w_1
            \sum_{n=0}^5
            \hat C^n_6
            \left(\ket{3_i}+\re^{\ri\varphi_i}\ket{3'_i}\right)
        \right)
        +
        \dots
    \nonumber\\
    &\ket{\Psi_{\rm II}} = \hat R_{\overline0}
        \left(
            w_0
            \ket{1_0}
            +
            w_1
            \sum_{n=0}^5
            \hat C^n_6
            \ket{1_1}
        \right)
        +
        \dots
    \label{eq:skyrmion_wave_function}
\end{align}
where we retained the classical magnetic configuration $w_{0}$ as well as up to twelve leading quantum corrections $w_k$ with parameter-dependent weights $w_1 = |w_{k}|$ which can differ by a relative phase we denote $\varphi_i$.
$\hat C_6$ denotes the operator associated with a $\pi/3$ rotation of the Fock state~\cite{SM}.
The notation $\ket{1_0}$ indicates a state with a single spin down at the center site, $\ket{1_1}$ a state with a spin down in the first shell surrounding the center, whereas $\ket{3_i}$ and $\ket{3'_i}$ are states with three down-spins distributed around and including the center site in four different configurations.
The regimes where these different projections contribute to the ground state are presented in \cref{fig:sub_regions}.

Notably, the rotated Fock state expansion reveals that the dominant quantum corrections involve only an odd number of down-spins, as visible in the $N_\downarrow=\{1,3\}$ regions of panel $(f)$.
More interestingly, the leading quantum corrections can take two qualitatively different forms:
The leading quantum corrections in $\ket{\Psi_{{\rm I},i}}$ are generated from $N_{\downarrow}=3$ states, whereby in $\ket{\Psi_{\rm II}}$, the leading corrections stem from $N_\downarrow=1$ states.
Therefore, we conclude that the skyrmion phase hosts at least two different types of quantum skyrmions, which have the same classical order but differ in their quantum corrections.
Compared to the classical ordered state, the fluctuations in the subleading term of $\ket{\Psi_{{\rm I},i}}$ involve flipped spins surrounding the skyrmion center, whereas in $\ket{\Psi_{\rm II}}$ they include the spin at the skyrmion center and one of its neighboring spins.
In \cref{fig:state_diagram}, panels $(e)$ and $(f)$, we respectively plot $|w_0|$ and $|w_1|$ together with the associated sum of down-spins $N_\downarrow = \sum_j n_{k,j}$.
Contour lines mark the domains of different $N_\downarrow$ sectors, where $w_0$ with $N_\downarrow=1$ extends over the full skyrmion region in the state diagram.
Given that the results reveal only comparatively small quantum fluctuations around an ordered state, on the order of $|w_1/w_0|\sim 1-10\%$ (see \cref{fig:state_diagram}, panels $(e)$ and $(f)$), it would be interesting to compare the results presented here with the outcomes of spin wave theory calculations, in particular magnon squeezed states~\cite{Kamra2019,Maeland2022_1,Maeland2022_2}.

Finally, we want to highlight that our Fock state expansion could be used as a variational ansatz for the wave function of semiclassical quantum skyrmions, which potentially leads to performant numerical optimization algorithms. 
The classical magnetic order is captured by $2$ rotation angles per site, resulting in $38$ parameters for the $19$-site quantum flake that, for the initial state, can be chosen to minimize the classical energy, i.e.,
\begin{align}
    E_c = \min_{\bm\theta,\bm\phi}\braket{\Psi_c(\bm\theta,\bm\phi) | \hat H | \Psi_c(\bm\theta,\bm\phi)}.
    \label{eq:classical_energy}
\end{align}
In addition to the local order, $\chi$ complex weights represent the quantum fluctuations.
In the semiclassical skyrmion regime, we recognize from \cref{eq:skyrmion_wave_function} and \cref{fig:state_diagram} panels $(e)$ and $(f)$ that, by using crystalline symmetries, only two different weights are needed to approximate the quantum mechanical ground state with high fidelity, as the decrease of $w_0$ is compensated by a simultaneous increase of $w_1$.
The simulation could then be performed by an independent iteration between the optimization of the angles and the weights until convergence is reached.
For comparison, we can estimate the number of variational parameters of an unconstrained matrix product state (MPS) simulation.
We find that comparable wave function fidelities are reached with bond dimension $M=3$, which corresponds to MPSs with $302$ complex variational parameters.
We checked that truncating the Fock state expansion leads to a more efficient representation for bigger skyrmion ground states and higher fidelities as well~\cite{SM}.

{\it Summary and Conclusion.--}
In this paper, we introduced a variational ansatz to represent quantum skyrmions as bosonic operators.
The form of this operator is determined by the desire to represent faithfully two fundamentally different aspects of a quantum skyrmions: the classical magnetic order and a ``quantum cloud'' of local spin-flip excitations around the classical magnetic order.
Using a minimal model for a chiral magnet in two dimensions, we found two distinct regions in the single-skyrmion phase: one where the leading quantum corrections around the classical magnetic order contain a flip of the center spin and one where they only contain spin flips around the skyrmion center.
We argued that the Fock basis of spin flips around the classical order can be particularly useful for numerical algorithms based on the variational principle, because limiting the allowed number of spin flips truncates the dimension of the numerical Fock space and thus reduces the amount of free parameters in the simulations.
We expect that such approaches can be fruitful when applied to or combined with tensor network simulations.
For this reason, it will be very interesting to investigate quantum skyrmions stabilized by antiferromagnetic frustration with similar techniques as the ones presented in this work.
Our work therefore paves the way toward a coarse-grained bosonic description of many-skyrmion quantum phases such as quantum skyrmion liquids which, unlike other approaches, incorporates individual skyrmions' microscopic quantum fluctuations.

{\it Acknowledgements.--}
We thank Alexander Mook, Thore Posske, Matteo Rizzi, Karin Everschor-Sitte and Niklas Tausendpfund for stimulating discussions.
T.L.S. and A.H. acknowledge financial support from the National Research Fund Luxembourg under Grants No. CORE C20/MS/14764976/TopRel and No. INTER/17549827/AndMTI. S.D. and W.B. acknowledge funding by the Excellence Strategy of the University of Konstanz via a Blue Sky project and by the Deutsche Forschungsgemeinschaft (DFG, German Research Foundation) via the Collaborative Research Center SFB 1432 project no.~425217212 and SPP2244 project no.~417034116.

\bibliography{biblio}
\clearpage\newpage
\onecolumngrid
\begin{center}
    \textbf{\large Supplemental Material for ``Quantum Magnetic Skyrmion Operator''} 
\end{center}
\setcounter{equation}{0}
\setcounter{figure}{0}
\setcounter{table}{0}

\makeatletter
\renewcommand{\theequation}{S\arabic{equation}}
\renewcommand{\thefigure}{S\arabic{figure}}
\renewcommand{\thetable}{S\Roman{table}}
\newcommand{\subf}[2]{%
  {\small\begin{tabular}[t]{@{}c@{}}
  #1\\#2
  \end{tabular}}%
}

\section{Rotated Fock states}
\label{sec:rotated_fock_states}
In this section, we construct a suitable orthonormal basis to expand the quantum skyrmion states with a semi-classical spin profile.
A classical skyrmion without quantum fluctuations can be written as a product of spin coherent states~\cite{Perelomov_book}
\begin{align}
    \ket{\Psi_c} = \hat R(\bm\theta,\bm\phi)\ket{\Uparrow}
\end{align}
where $\hat R$ is a product of local spin rotations, denoted by the polar $\bm\theta$ and azimuth angles $\bm\phi$, and $\ket{\Uparrow}$ represents the fully polarized spin state, i.e.
\begin{align}
    \ket{\Uparrow} = \bigotimes_i\ket{\uparrow}_i
    ,\ 
    \hat R(\bm\theta,\bm\phi) = \prod_i \re^{-\ri \phi_i \hat S_{i,z}}\re^{-\ri \theta_i \hat S_{i,y}}.
\end{align}
We find that the best classical approximation to the quantum skyrmion state is obtained by a rotation operator which redirects the normalized spins of the ordered product state along the direction of the spin expectation values of the quantum state.
We denote this product state as $\ket{\Psi_c}$, which is equivalent to $\ket{\Uparrow}$ up to local spin rotations by definition.
Note that in this convention, $\ket{\Psi_c}$ corresponds to a mean-field approximation of the ground state where all quantum fluctuations are neglected.
Depending on the amount of quantum entanglement in the ground states $\ket{\Psi}$, we find overlaps $|w_0| \coloneqq |\braket{\Psi_c|\Psi}| \geq 70.17\%$, presented in \cref{fig:state_diagram} $(e)$.
The polarized states found in the bottom left region are equivalent to product states $\ket{\Uparrow}$ with $|w_0|=1$, whereas quantum skyrmion ground states with finite entanglement display $|w_0| < 1$.
Large quantum fluctuations in the ground state result in magnetic textures without classical local order, which is indicated by \cref{fig:state_diagram} panels $(a)$ and $(c)$ in the pillar-shaped region where the norm of the spin expectation value is vanishing $|\braket{\hat{\bm S}_i}|\approx 0$ and the Rényi entropy $H_{2,0}\approx\ln2$.
With exact diagonalization, it is hard to verify whether the eigenstates in this region assume a finite spin norm in larger systems.
However, a limited finite-size extrapolation (different sizes and system shapes) indicates a vanishing spin norm in the thermodynamic limit.
The overlap to a product state must vanish exactly whenever $|\braket{\hat{\bm S}_i}|\rightarrow0$, marking a clear indicator when the employed Ansatz of \cref{eq:quantum_skyrmion_operator} fails.

Quantum effects in the magnetically ordered ground states can be taken into account by expanding the quantum wave function in the basis of locally spin-flipped states, i.e.
\begin{align}
    \ket{\bm n} = \prod_i \left(\hat S^-_i\right)^{n_i}\ket{\Uparrow} 
\end{align}
where $\bm n\in \{0,1\}^{\otimes N}$ is a vector associated with the total number and position of down spins.
The Fock basis $\{\ket{{\bm n}_i}\}$ spans a complete basis for the Hilbert space associated with $\hat H$.
Note that the central spin of a skyrmion is anti-aligned with the polarized environment, and surrounded by continuous rotations of spins -- as a result, the overlap $\braket{{\bm n}_i|\Psi}$ is small for all Fock states.

\begin{figure}[ht]
    \includegraphics{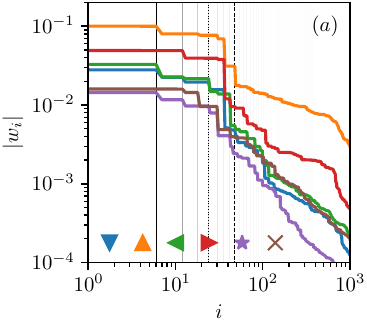}
    \hfil
    \includegraphics{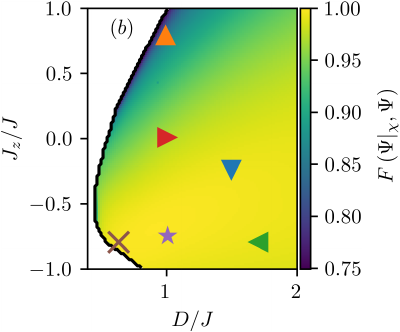}
    \caption{
    In panel $(a)$ we show the decay of the amplitudes $|w_i|$ of the corrections to the classical skyrmion state.
    Weights occur in multiples of $6$, marked by grey grid lines, which motivates using the rotation operator $\hat C_6$ associated with this symmetry.
    In panel $(b)$, we show the fidelity $F(\Psi|_{\chi}, \Psi)$ when we retain $\chi=6$ corrections in addition to the classical skyrmion state $|w_0|\geq 70.17\%$ and mark the parameters of the ground states associated with the decay of amplitudes presented in $(a)$ with different symbols.
    The dotted (dashed) line in panel $(a)$ marks $\chi=24\,(48)$, for which the state fidelity reaches $F(\Psi|_{\chi}, \Psi) = \sqrt{\sum_{i=0}^\chi |w_i|^2} \geq 90 \,(95)\%$.
    Since amplitudes decay quickly in magnitude, the skyrmion creation operator of \cref{eq:quantum_skyrmion_operator} can be an efficient approximation when weights smaller than a threshold value are discarded in the sum.}
    \label{fig:weights_decay}
\end{figure}

This issue is overcome by using a {\it rotated} Fock basis where the local spin flips are aligned with the canonical axis of the magnetic order $\{\hat R\ket{{\bm n}_i}\}$.
In this basis, the Fock state with the highest overlap to the true ground state $w_0 \coloneqq \braket{\Psi_c|\Psi}$ is given by $\ket{\Psi_c} = \hat R\ket{\Uparrow}$.
We assume that the lattice is commensurate with the skyrmion center at site ``0'', i.e., the discretized skyrmion profile is spin-flipped at a central lattice site concerning the polarized environment.
In this case, we can conveniently associate a lattice site with the center-of-mass of the skyrmion, which is highlighted more explicitly by replacing the $\pi$-rotation in $\hat R$ by a spin ladder operator, i.e. $\hat R \rightarrow \hat R_{\overline0}\hat S^-_{0}$.
We now expand a state $\ket\Psi$ in the rotated Fock basis, which can be ordered by descending magnitude $|w_i|\geq|w_{i+1}|$ 
\begin{align}
    \ket\Psi = \frac{\hat R_{\overline0}}{\sqrt{\sum_i^{\chi'}|w_i|^2}} \sum_{i=0}^{\chi'} w_i \ket{\bm n_i}.
    \label{eq:wavefunction_fock_expansion}
\end{align}
With the convention of $\hat R_{\overline 0}$, we exchanged the state associated with the leading contribution from $\ket\Uparrow$ to  $\ket{\bm n_0}$ which retains a spin down at the center site, i.e. $n_{0,0}=1$.
$\chi'$ denotes the number of states contributing to the wave function, and in principle is growing exponentially in the number of sites, i.e. $\chi'=2^N$.
In this case, the sum of squared weights in the denominator of \cref{eq:wavefunction_fock_expansion} is $1$ for normalized states.
We can, however, attempt to approximate any state by truncating the sum to a smaller $\chi<\chi'$ while retaining a reasonable state fidelity.
As described in the main text, we find a large regime in the state diagram that hosts quantum skyrmions (see \cref{fig:sub_regions}), which is divided into two qualitatively different quantum skyrmion flavors, described by $\ket{\Psi_{{\rm I},i}}$ and $\ket{\Psi_{\rm II}}$.
In these wave functions, we noticed that Fock states equal up to six-fold rotations contribute with the same absolute magnitude to the wave functions, such that we use a rotation operator $\hat C_6$ to encode this symmetry.
The action of the rotation operator $\hat C_6$ on the spin-flipped Fock states is given by $\hat C_6^6=\hat C_6^0=1$ and 
\begin{align}
    \hat C_6^5 \Ket{\includegraphics[valign=B,raise=-0.55cm,width=0.6cm]{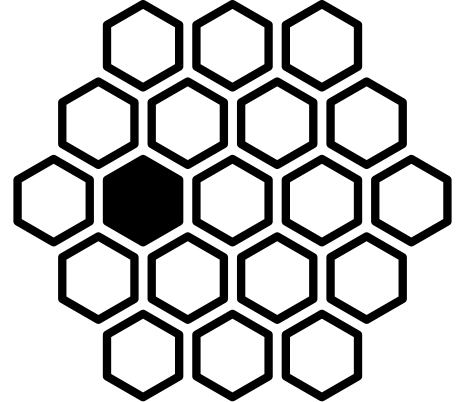}}
    =
    \re^{-\ri\pi/3}\hat C_6^4\Ket{\includegraphics[valign=B,raise=-0.55cm,width=0.6cm]{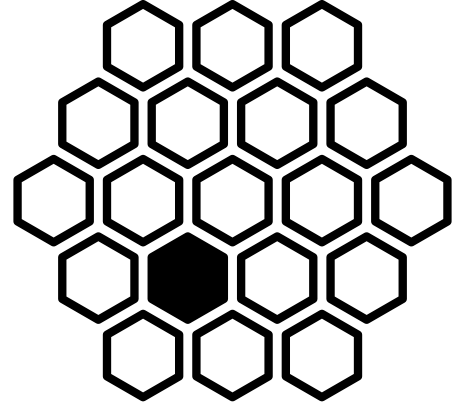}}
    =
    \re^{-2\ri\pi/3}\hat C_6^3\Ket{\includegraphics[valign=B,raise=-0.55cm,width=0.6cm]{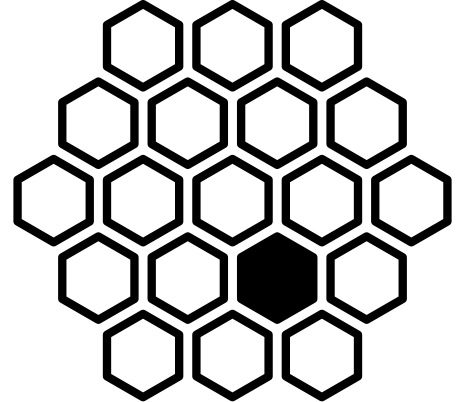}}
    =
    \re^{-\ri\pi}\hat C_6^2\Ket{\includegraphics[valign=B,raise=-0.55cm,width=0.6cm]{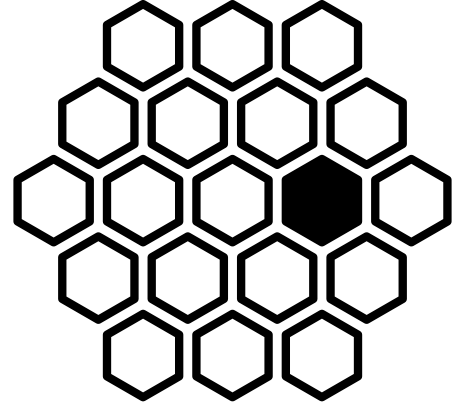}}
    =
    \re^{2\ri\pi/3}\hat C_6\Ket{\includegraphics[valign=B,raise=-0.55cm,width=0.6cm]{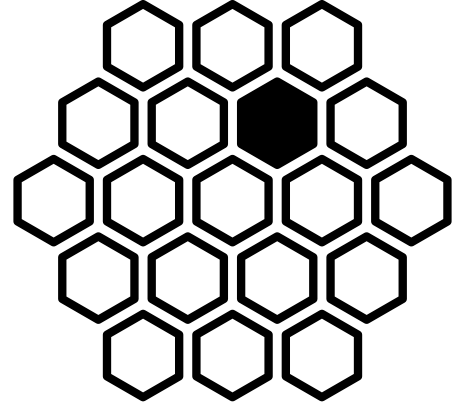}}
    =
    \re^{\ri\pi/3}\Ket{\includegraphics[valign=B,raise=-0.55cm,width=0.6cm]{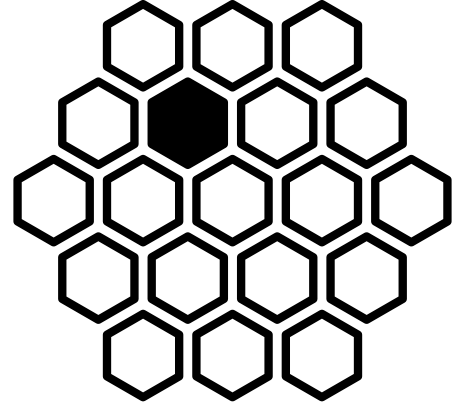}}
    .
\end{align}
We find that the leading $N_\downarrow=3$ states presented in \cref{fig:sub_regions} rotate under $\hat C_6$ in a straightforward way without additional phases.
We now define $\ket{1_0}=\ket{\includegraphics[valign=B,raise=-0.45cm,width=0.45cm]{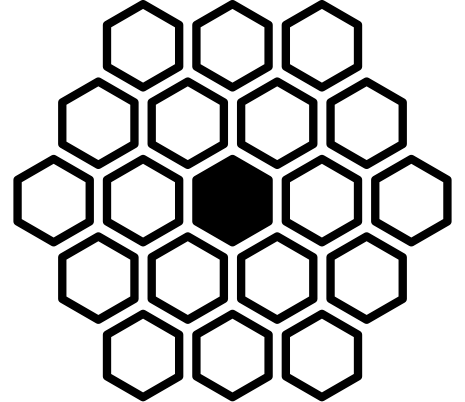}}$ and $\ket{1_1}=\ket{\includegraphics[valign=B,raise=-0.45cm,width=0.45cm]{3.png}}$, as well as the $\ket{3_i}$ states of \cref{fig:sub_regions} for brevity, and find the expression for the ground states presented in \cref{eq:skyrmion_wave_function}, which corresponds to the leading contributions of \cref{eq:wavefunction_fock_expansion}.
To estimate the efficiency of the employed Ansatz of the skyrmion operator, we show in \cref{fig:weights_decay} that the different amplitudes $|w_i|$ decay quickly, with increasing $i$, in most parts of the phase diagram.

\section{Observables}
\label{sec:observables}
To evaluate the reliability of the above approximation, we make use of several observables, such as the fidelity between two arbitrary states $\ket\Psi$ and $\ket\Phi$
\begin{align}
    F(\Psi, \Phi) = |\Braket{\Psi|\Phi}|
\end{align}
which will be the most reliable quantity to estimate the quality of an approximation.
In particular, the fidelity of the approximation when we truncate the sum in \cref{eq:wavefunction_fock_expansion} is equal to the squared sum of weights $F(\Psi|_{\chi}, \Psi) = \sum_{i=0}^{\chi}|w_i|^2 \leq 1$.
In \cref{fig:weights_decay} panel $(b)$, we see that a reasonable state fidelity $\geq75\%$ is reached by keeping only the first six ($\chi=6$) quantum corrections.
In panel $(a)$, the dotted and dashed lines at $\chi=24$ and $\chi=48$ indicate when the fidelity reaches $90\%$ and $95\%$, respectively.

To investigate the magnetic texture, we compute the local spin expectation values
\begin{align}
    \bm S_i = \Braket{\hat{\bm S}_i}
\end{align}
which, if $|\bm S_i|\approx s$, indicate magnetically ordered states.
We further characterize magnetic order by the integer-valued ``topological charge'' $q$ defined in terms of the normalized spin expectation values $\bm O_i = \bm S_i / |\bm S_i|$ as
\begin{align}
    q &= \frac1{2\pi}\sum_{\braket{ijk}}\arg\left(\frac{X_{ijk}+\ri Y_{ijk}}{\sqrt{X_{ijk}^2 + Y_{ijk}^2}}\right),
    \\
    X_{ijk} &= 1 + \bm O_i\cdot \bm O_j + \bm O_j\cdot \bm O_k + \bm O_k\cdot \bm O_i,
    \\
    Y_{ijk} &= \bm O_i\cdot (\bm O_j\times \bm O_k),
\end{align}
where $\braket{ijk}$ denotes a sum over elementary triangles of the lattice~\cite{Berg1981}.
A vanishing $q$ indicates topologically trivial magnetic textures as with the field-polarized and other collinear states, while a non-zero $q$ reveals topological magnetic textures such as magnetic skyrmions.

\section{Adiabatic exchange of two quantum skyrmions with MPS}
\label{sec:MPS_braiding}
One characteristic quantity related to the elementary form of a quantum mechanical many-body wave function is the statistical phase after particle exchange.
In particular, a wave function of two bosons (fermions) $\Psi(\bm x_1,\bm x_2)$, with $\bm x_1\neq \bm x_2$ the different positions of the two identical particles, acquires a statistical phase $+1$ ($-1$) after particle exchange $\Psi(\bm x_1,\bm x_2)\rightarrow \Psi(\bm x_2,\bm x_1)=\pm\Psi(\bm x_1,\bm x_2)$.
A geometric way to obtain the exchange phase is through the Berry phase~\cite{Berry1984,Nayak2008}.
To compute the Berry phase, we add to the Hamiltonian in \cref{eq:hamiltonian} of the main text the following pinning potential
\begin{align}
    \hat V(\lambda) = V_0\sum_{j=1}^N\sum_{i=1}^2\exp\left(-\frac{\left|\bm R_j - \bm x_i(\lambda)\right|^2}{2\sigma^2}\right)\hat S_{z,j} \,.
\end{align}
In the resulting Hamiltonian $\hat H(\lambda)$, if the exchange couplings are tuned such that the ground state is polarized to the external field, using $V_0\gg J$ pins exactly two semi-classical skyrmion profiles around the center-of-mass positions $\bm x_1(\lambda)$ and $\bm x_2(\lambda)$.
Smoothly deforming the pinning potential so that $\bm d(\lambda)\coloneqq \bm x_2(\lambda) - \bm x_1(\lambda)$ rotates by $\pi$ in the interval $\lambda\in[0,1]$, defines a closed path in the parameter space of $\hat H(\lambda)$ where the two identical skyrmions exchange positions.
If $\lambda(t)$ is changed adiabatically in time, we can assume that the ground state $\ket{\Psi}$ follows the parameter deformation instantaneously, such that $\ket{\Psi(t)}=\ket{\Psi(\lambda(t))}$.

\begin{figure}[ht]
    \centering
    \includegraphics[valign=c]{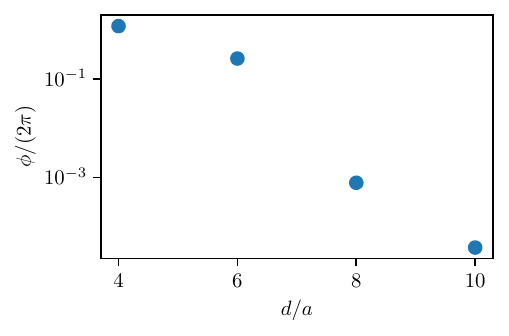}
    \hfil
    \begin{tabular}{l}
    \subf{\includegraphics[width=40mm]{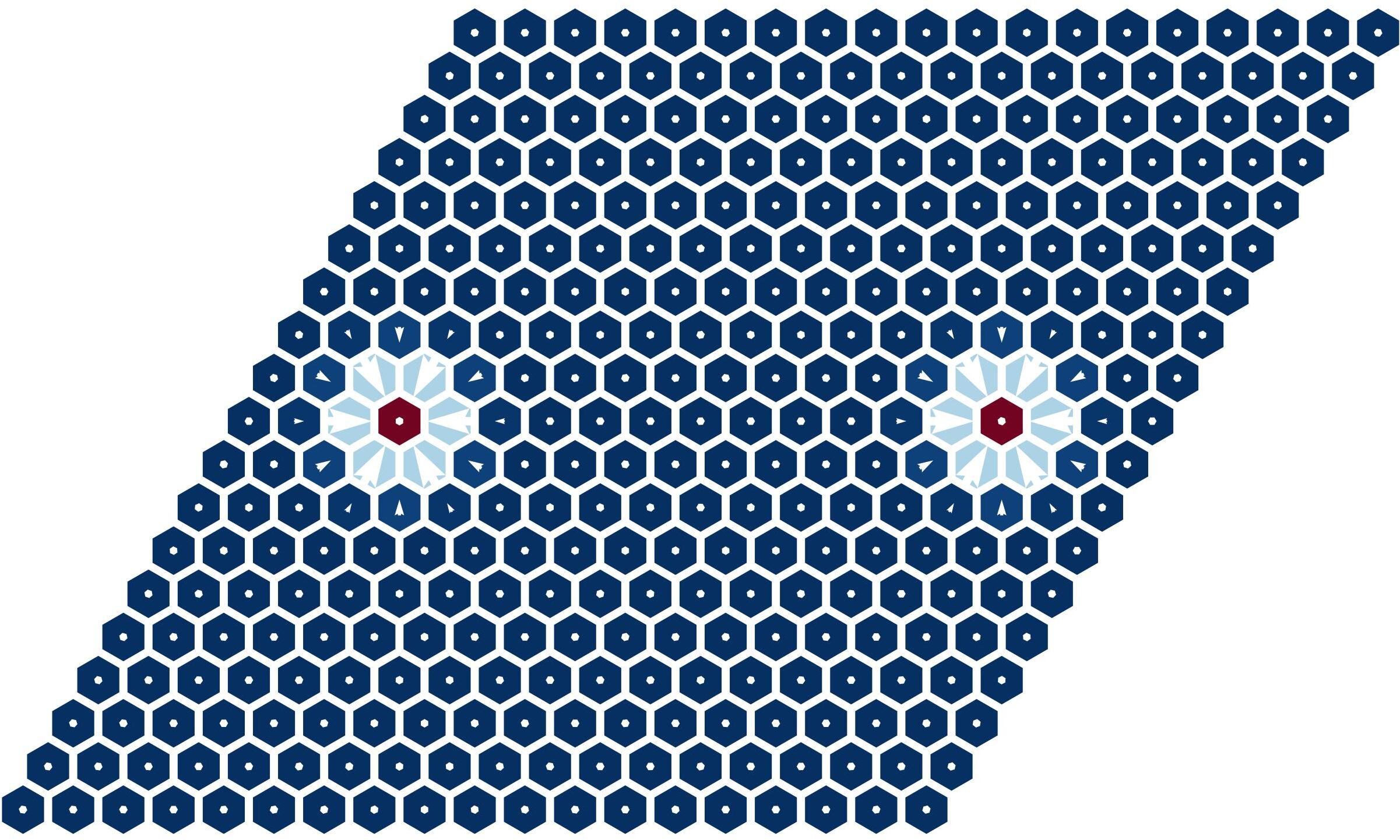}}{Initial two-skyrmion configuration}
    \\
    \subf{\includegraphics[width=40mm]{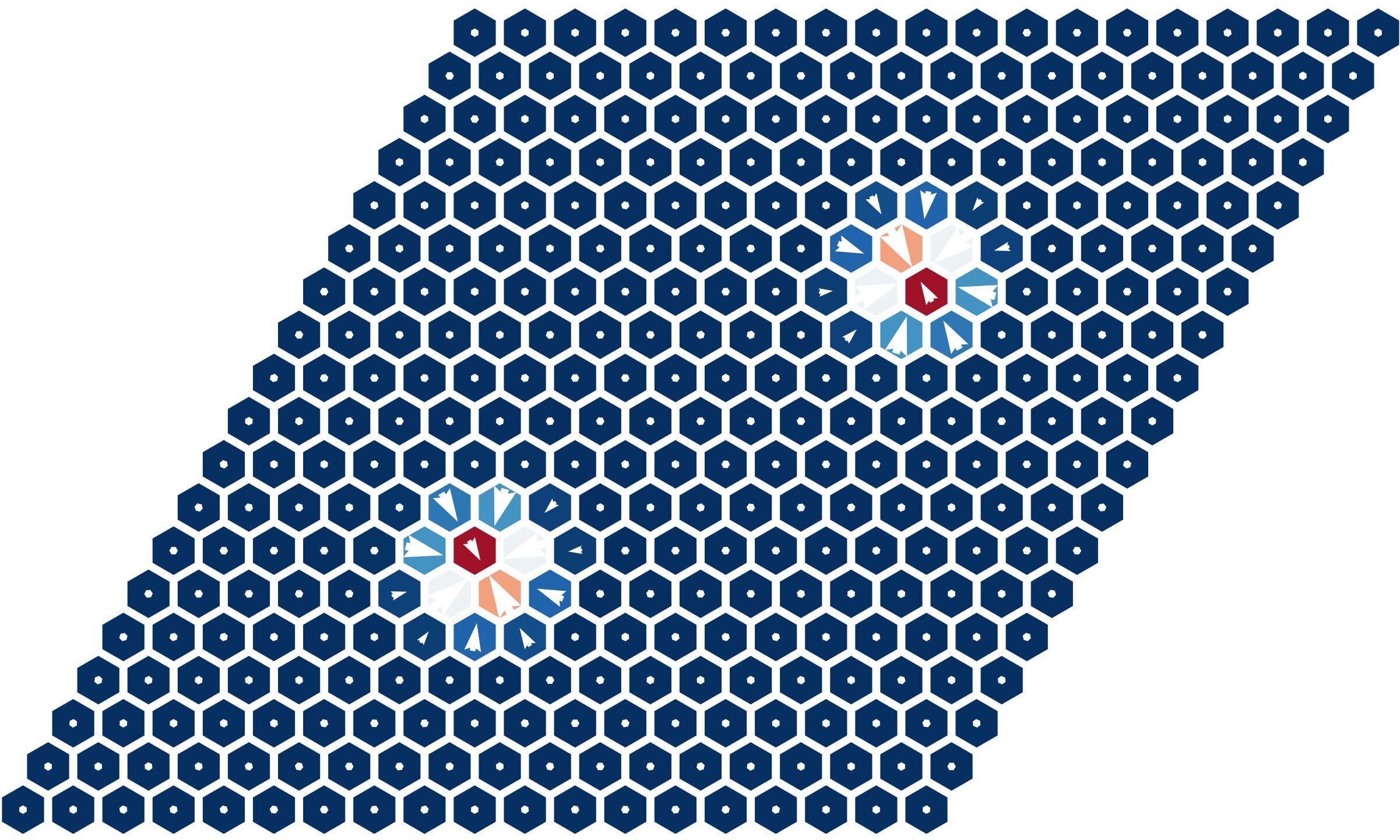}}{After $1/8$ of the braiding cycle.}
    \end{tabular}
    \caption{Exchange angle versus distance for a semi-classical magnetization profile with two skyrmions using a strong pinning field $V_0=4J$, $\sigma=a$. We recognize an exponentially vanishing trend in the distance $d$, compatible with an approximately bosonic two-skyrmion wave function. Snapshots during the exchange process are depicted on the right, where the color represents $z$- and arrows the $x/y$-components of the magnetization.}
    \label{fig:exchange_angle}
\end{figure}

For skyrmion profiles with radius $r_c\approx 3a$, as in the main text, a well-separated two-skyrmion magnetization profile already requires system sizes that cannot be investigated by exact diagonalization.
Instead, we proceed by using MPS simulations for larger quantum systems, which are known to faithfully capture the properties of quantum skyrmion lattice ground states with modest computational cost~\cite{Haller2022}.
The Berry phase is then extracted by (i) using variational MPS to evaluate the ground states of a discrete sequence in the adiabatic deformation, i.e. $\{\ket{\Psi_n}\coloneqq\ket{\Psi(n\Delta\lambda)}, n=0,...,N_\lambda-1\}$ with $\Delta\lambda = 1/N_\lambda$, and (ii) evaluating the Berry phase through a Wilson loop
\begin{align}
    \phi = \arg\prod_n \frac{\braket{\Psi_n|\Psi_{n+1}}}{|\braket{\Psi_n|\Psi_{n+1}}|}
    .
\end{align}
It is easy to recognize that the Berry phase $\phi$ obtained by the Wilson loop is invariant under gauge transformations $\ket{\Psi_n}\rightarrow\re^{\ri\varphi_n}\ket{\Psi_n}$ by definition, such that we do not need to fix a gauge for the numeric simulations.
The exchange angle (Berry phase) $\phi$ numerically computed at various skyrmions separations $d$ is shown in \cref{fig:exchange_angle}.
For $d$ larger than the skyrmion radius $r_c$, the exchange angle exhibits an exponentially vanishing trend in $d$.
Since for bosons ${\rm mod}_{2\pi}\phi=0$, it is reasonable to argue that the two skyrmions behave as bosonic quasiparticles if they are separated over large distances.
On short distances comparable with $r_c$, the exchange phase is significantly larger, which is also expected because the rotation operators associated with the classical magnetic order overlap~\cite{Istomin2000}.
Our numerical results therefore agree with the hard-core bosonic nature of the quantum skyrmion operator.

\section{Scaling of the quantum fluctuations with the spin}
\label{sec:spin_scaling}
\begin{figure}[ht!]
    \centering
    \includegraphics{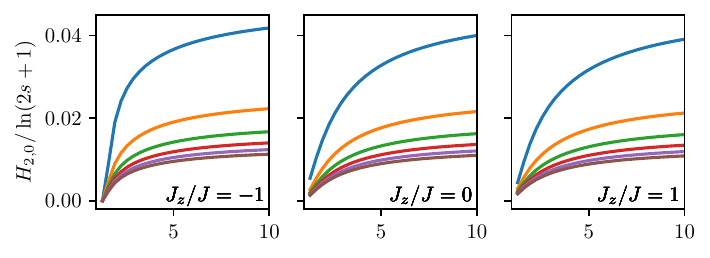}
    \includegraphics{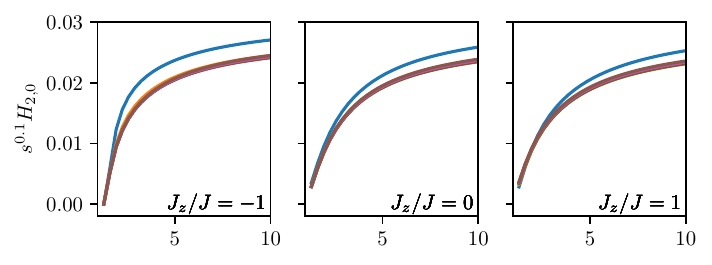}
    \includegraphics{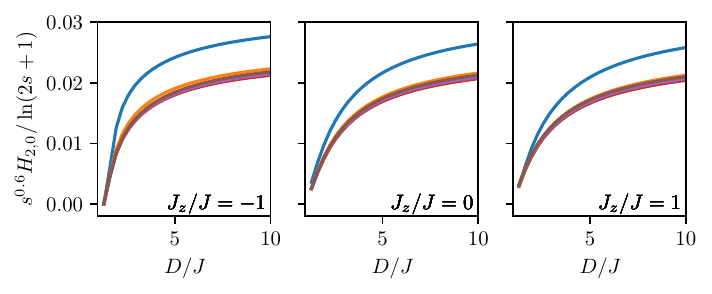}
    \includegraphics{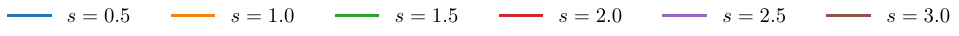}
    \caption{The system's central Rényi entanglement decreases with the value of the spin $s$ at different parameters $J_z/J$ and $D/J$. Here, we focus strictly on the quantum skyrmions regime of the state diagram. We recognize that the scaling of the (relative) entanglement with the spin is compatible with $s^{-0.1}$ ($s^{-0.6}$) and therefore to a classical limit with vanishing entanglement.}
    \label{fig:spin_scaling}
\end{figure}

To show that the results can extend to spin values $s>1/2$, we performed additional simulations on a small $N=7$ site spin-$s$ cluster ($6$ spins surrounding the center in a triangular configuration), and present the results in \cref{fig:spin_scaling}.
Note that the results in \cref{fig:state_diagram} mostly concern the $N=19$ site cluster, for which memory constraints allow us to investigate only small spin values $s=1/2$.
However, the states found in smaller clusters are qualitatively similar, with the benefit that $N=7$ site systems allow us to investigate cases $s\leq 3$, with a comparable computational cost as $N=19$ and $s=1/2$ flakes.

To investigate the entanglement in the system, we computed the second Rényi entropy $H_{2,0}$ of the center spin.
The maximum value of the entropy is $\ln(2s+1)$, such that $H_{2,0}/\ln(2s+1)$ estimates the relative entanglement in the system accounting for the increasing dimension of the local Hilbert space.
Based on these results, we find a trend towards a ``classical limit'' with vanishing entanglement.

We would like to mention that the growth of the dimension of the local Hilbert space $d=2s+1$ is only linear in $s$, whereas the dimension of the $N$-site Hilbert space grows exponentially as $d^N$.
The qualitative challenge of the exponential growth of the many-body Hilbert space is therefore independent of $s$ and tackled for the specific case $s=1/2$ in our present work.
We do not claim that the investigation of magnetic flakes with higher spin values as a hypothetical extension of our investigation is entirely trivial, but it is, in our opinion, straightforward to generalize:
In this case, $\hat S^\pm$ and $\hat S_\alpha$ are the spin-$s$ ladder and $\alpha\in\{x,y,z\}$ operators, and occupation numbers $n_{k,j}$ can assume values from $0,...,2s$.
Although there is no fundamental reason to expect a monotonic decrease of quantum fluctuations with increasing $s$, the physical intuition is that a scaling towards large spins approaches a classical limit with the absence of quantum fluctuations, such that the proposed variational wave function becomes an exact representation of the coherent state and coincides with an optimization of the spin angles.
This argument is compatible with the preliminary results from the scaling we found in \cref{fig:spin_scaling}, and we expect that the skyrmion operator formalism is thus applicable for general spin values.

\section{Scaling of the quantum fluctuations with the system size}
\begin{figure}[ht!]
    \centering
    \begin{tabular}{c c c c c}
        $N=37$ & $N=61$ & $N=91$ &  $N=127$\\
        \includegraphics[width=0.224\textwidth]{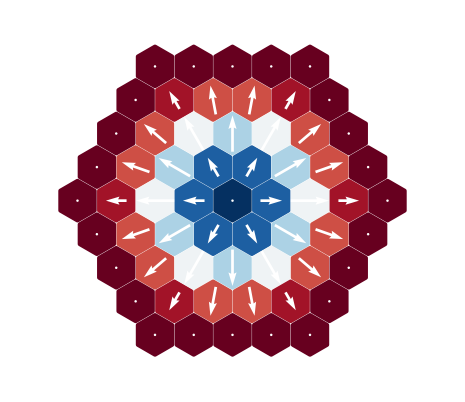} &
        \includegraphics[width=0.224\textwidth]{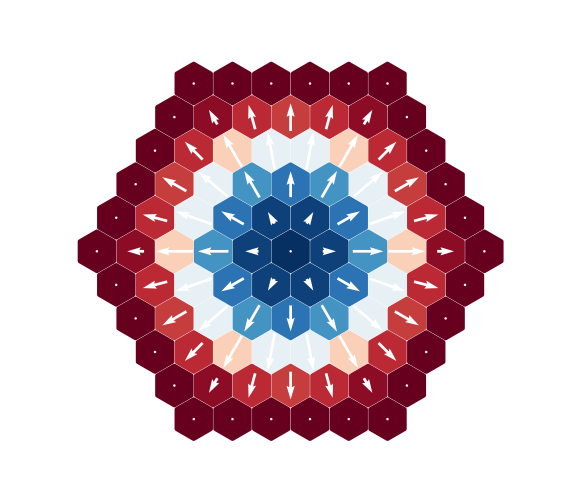} &
        \includegraphics[width=0.224\textwidth]{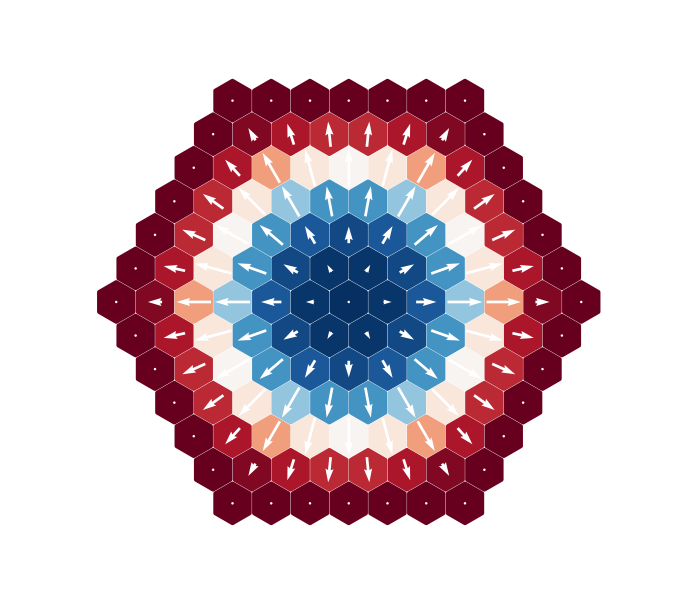} &
        \includegraphics[width=0.224\textwidth]{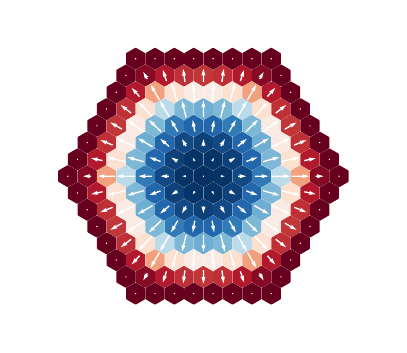} &
        \includegraphics[height=3.5cm]{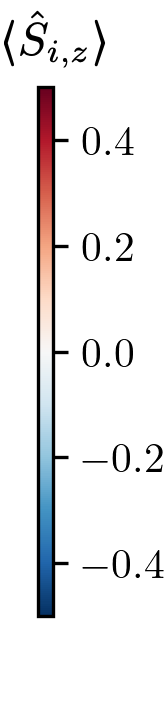} \\
        \includegraphics[width=0.224\textwidth]{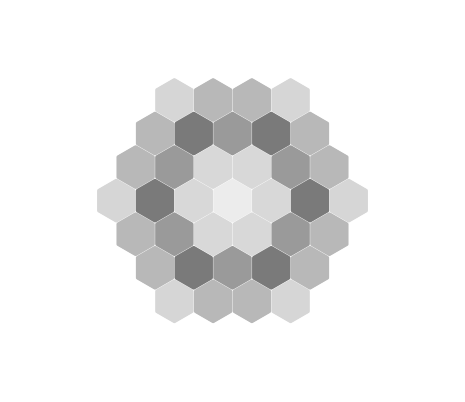} &
        \includegraphics[width=0.224\textwidth]{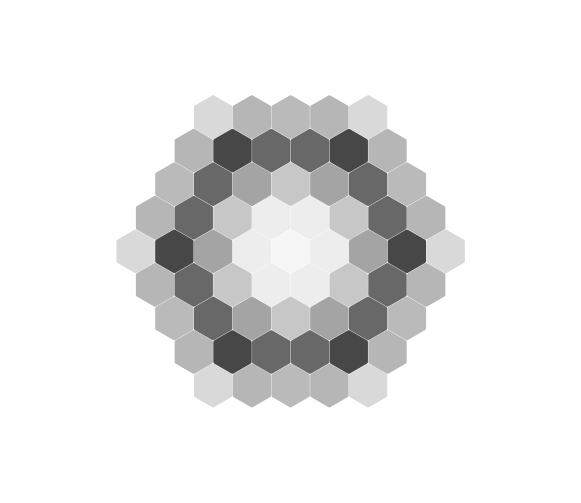} &
        \includegraphics[width=0.224\textwidth]{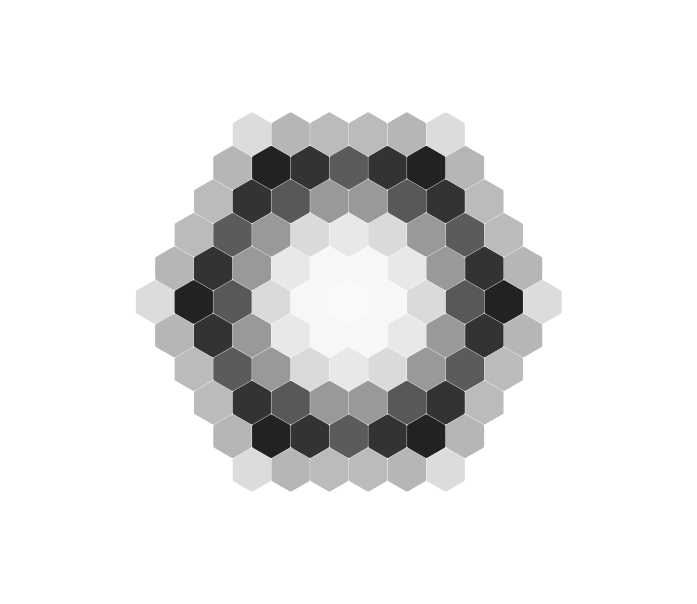} &
        \includegraphics[width=0.224\textwidth]{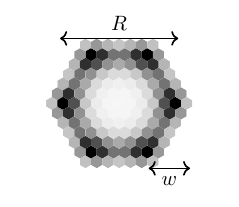} &
        \includegraphics[height=3.5cm]{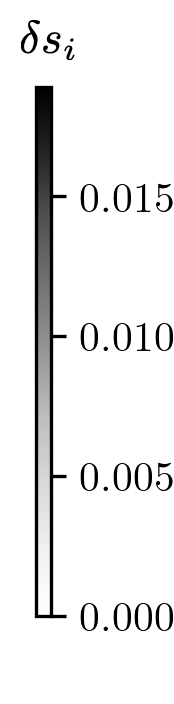} \\
    \end{tabular}
    \caption{Spin expectation values $\braket{\bm S_i}$ (arrows and color) and spin norm deviation $\delta s_i = 1 - |\braket{\bm S_i}|/s < 0.02$ (in grey). The blue-red color gradient visualizes $z$ (from $-s$ to $s$) and arrows depict $x,y$ components, whereas the linear black to white gradient represents $\delta s_i$. We recognize that $\delta s_i$ localizes around the skyrmion center, where spins have large in-plane $(x,y)$ components. Parameters are chosen as $B=0$, $J_z=1.13J_x=1.13J_y$, $J_x=J_y=-D$.}
    \label{fig:size_scaling}
\end{figure}

To investigate how the cutoff parameter $\chi$ should be chosen in relation to the relevant length scales of the skyrmions in the target system, we study the quantum fluctuations of larger flakes, i.e. up to $N=127$ sites (see \cref{fig:size_scaling}), obtained by DMRG simulations.
We obtained these results by fixing the bond dimension to $M=128$.
The magnetization of the skyrmion state, $\bm m(\bm r) = \braket{\hat{\bm S}_{\bm r}}$, with our convention of the DMI vector along $\bm e_z$, is close to a classical limit with
\begin{align}
    \bm m \approx s (\sin\Theta\cos\phi, \sin\Theta\sin\phi, \cos\Theta)
    ,\quad
    \Theta(r) = \pi - 2\arctan\left(\frac{\sinh(r/w)}{\sinh(R/w)}\right)
\end{align}
where $\phi$ is the azimuth angle of $\bm r$, $w$ describes the width of the domain surrounding the skyrmion core and $R$ is the radius of the skyrmion core.

The precise values of $R$ and $w$ depend on the microscopic coupling parameters \cite{Wang2018}.

Based on the results presented in \cref{fig:size_scaling}, we notice that the quantum fluctuations are sharply localized at the domain-wall surrounding the skyrmion core, signalled by the deviation of the spin norm from the quantized value $s$, i.e. $\delta s_i = 1 - |\braket{\bm S_i}|/s$.
In the region where $\delta s$ is sizeable, i.e. the black rings of \cref{fig:size_scaling} spanned by the diameter $2\pi R$ and the width $w$ of the domain-wall, spin-flip fluctuations are relevant corrections and should be captured by the skyrmion operator.
The signature in $\delta s$ is caused by localized domain-wall entanglement, which was reported in previous works, e.g.~\cite{Haller2022,Joshi2023}.

If we allow Fock states where the basis elements contain up to $N_\downarrow = k$ down spins, the Ansatz consists of $\chi=\sum_{j<k}\frac{N!}{j!(N-j)!}\sim N^k$ terms.
If we further assume that spin-flips are confined to the $N_{\rm dw}$ sites which form the domain-wall surrounding the skyrmion, and that $N_{\rm dw}$ scales with $R$ and $w$, we reduce the former to a polynomial scaling $\chi \sim (w R)^k$.

Many-skyrmion systems may form collective structures such as quantum skyrmion lattices~\cite{Haller2022}.
For quantum skyrmion lattices, the profiles of individual solitons may be slightly distorted, but the entanglement of quantum skyrmion lattices is still bound to the domain-wall of individual skyrmions and the collective state is approximately a product state of individual quantum skyrmions.
In these situations, the scaling of the variational parameters is polynomial in the skyrmion radius and linear in the number of skyrmions
\begin{align}
    \chi \sim N_{\rm sk} (wR)^k
    \label{eq:size_scaling}
    .
\end{align}
Further constraints such as locality between the down-spins in the Fock states, or conserving crystal symmetries (such as the $\hat C_6$ symmetry exploited in \cref{eq:skyrmion_wave_function}), could be imposed to arrive at a more favorable scaling without loss of accuracy.

The MPS Ansatz is composed of $N$ tensors, which are arrays of size $M_{i-1}\times M_i\times d$, where $M_i$ is called the bond dimension of the bond between sites $i$ and $i+1$.
The number of variational parameters in the MPS Ansatz at a fixed $M_i\leq M$ is the sum of all elements of the individual tensors and scales as $dM^2N$ for large systems.
However, the required $M$ scales with the amount of quantum fluctuations in the state, as the entanglement entropy of an MPS is bound by the bond dimension, i.e. $S=\ln M$ for the von-Neumann entanglement entropy.
General statements on the scaling of $M$ with the system size can be made for one-dimensional systems:
In the case of short-ranged interactions, it has been shown by Hastings~\cite{Hastings2007} that ground states follow the area-law of entanglement, which states that the entropy is $S\sim\ln\xi/a$.
To our knowledge, general proofs of an area law do not exist for two-dimensional spin systems, and it is not clear which bond dimension scaling is expected for the quantum skyrmion phase.

To estimate the needed variational parameters in bigger systems, we therefore proceed as follows:
Given the MPS ground states $\ket{\Psi_M}$ of $M=128$, we approximate the state with the variational skyrmion Ansatz based on \cref{eq:wavefunction_fock_expansion}
\begin{align}
    \ket{\Psi_{\rm sk}} = \frac{\hat R_{\overline0}}{\sqrt{\sum_i^{\chi}|w_i|^2}} \sum_{i=0}^{\chi} w_i \ket{\bm n_i}
\end{align}
by keeping Fock states $\ket{\bm n_i}$ with up to three occupied down spins $\sum_j n_{i,j} \leq 3$ for all $i$ that contribute more than $|w_i|\geq 10^{-3}$.
In this approximation $\chi\ll 2^N$, which corresponds to the number of variational parameters to capture the quantum fluctuations.
Finally, we truncate the MPS bond dimension such that compatible fidelities are reached and compare the number of non-vanishing complex-valued variational parameters in \cref{tab:parameters_comparison}.
It demonstrates that our proposed variational skyrmion wave function, when compared to MPS, contains a smaller number of parameters.

\begin{table}[ht]
    \centering
    \begin{tabular}{ | c | c c | c c | c | }
        \hline\hline
        $N$ & $F_{\rm MPS}$ & $\#_{\rm MPS}$ & $F_{\rm sk}$ & $\#_{\rm VS}$ & $|w_0|$ \\
        \hline
         19 & 0.9994 &  762 & 0.9995 &  161 & 0.9850\\
         37 & 0.9980 & 1660 & 0.9984 &  467 & 0.9667\\
         61 & 0.9940 & 2858 & 0.9954 &  953 & 0.9391\\
         91 & 0.9859 & 4355 & 0.9892 & 1607 & 0.9037\\
        127 & 0.9747 & 6122 & 0.9785 & 2357 & 0.8613\\
        \hline\hline
    \end{tabular}
    \hfil
    \includegraphics[valign=c,height=2cm]{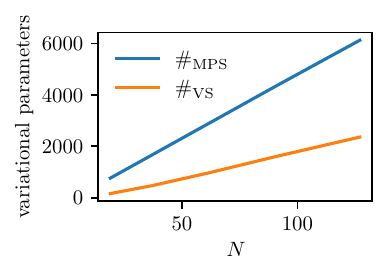}
    \caption{Comparison of the number of nonzero complex-valued variational parameters $\#_{\rm MPS/VS}$ in the variational skyrmion wave function and unconstrained MPS $\ket{\Psi_M}$ with compatible fidelities, which is achieved with bond dimension $M=5$. The reference state is given by the MPS state $\ket{\Psi_{128}}$ with bond dimension $128$ and fidelities are computed as $F_{\rm MPS}=\braket{\Psi_{128}|\Psi_{5}}$ and $F_{\rm sk}=\braket{\Psi_{128}|\Psi_{\rm sk}}$. Coupling parameters are $B=0$, $J_z=1.13J_x=1.13J_y$, $J_x=J_y=-D$.
    The values $|w_0|$ illustrate decreasing overlaps between the quantum state and the classical coherent state when we increase the system size.
    Note that $\#_{\rm VS}=\chi+N$ includes the $2N$ real-valued angles.
    Our proposed variational skyrmion wave function, when compared to MPS, contains a smaller number of parameters.
    }
    \label{tab:parameters_comparison}
\end{table}
\end{document}